\documentclass[preprint,showpacs,preprintnumbers,amsmath,amssymb]{revtex4}
\usepackage{amsmath,amssymb,latexsym,epsfig,graphicx,epstopdf,epsf,pdfpages}
\begin{document}
\title{Electron states and quantum transitions in a quantum ring on a sphere}
\author{Eduard M Kazaryan${}^1$, Vanik A Shahnazaryan${}^1$,  Hayk A Sarkisyan${}^{1,2}$}
\altaffiliation{email:shayk@ysu.am}
\affiliation{${}^1$Russian-Armenian (Slavonic) University, 0051 Yerevan, Armenia}
\affiliation{${}^2$Yerevan State University, 0025 Yerevan, Armenia}

\begin{abstract}
An analytical solution of the quantum problem of an electron on a spherical segment with angular confinement potential of the form of rectangular impenetrable walls is presented. It is shown that the problem is reduced to finding solution of hypergeometric equation. As an application of the obtained results the quantum transitions in this system are discussed, and it is shown that the selection rule for quantum number $l$ is removed due to the violation of spherical symmetry of the problem.


\end{abstract}

\maketitle

\section{introduction}

Investigation of quantum mechanical problems on curved surfaces initially had purely academic character and were perceived as interesting model problems. The exactly solvable problems of particle states on curved surfaces of different symmetry were discussed [1-8]. In particular, in Ref. [1] path integral formulations for the Smorodinsky -Winternitz potentials in two- and three dimensional Euclidean space are presented. In Ref. [2] path integral formulations for Smorodinsky -Winternitz potentials, respectively systems with accidental degeneracies, on the two- and three-dimensional sphere, and a complete classification of super-integrable systems on spaces of constant curvature are presented. All coordinate systems which separate the Smorodinsky-Winternitz potentials on a sphere, and state the corresponding path integral formulations are mentioned. In Ref. [3] the basis functions for classical and quantum mechanical systems on the two-dimensional hyperboloid that admit separation of variables in at least two coordinate systems are examined. In Ref. [4] the free quantum motion on the three-dimensional sphere in ellipso - cylindrical coordinates is studied, where à distinction between prolate elliptic and oblate elliptic coordinates is made. In Ref. [5] generalizations to spheres of Levi-Civita, Kustaanheimo-Steifel and Hurwitz regularizing transformations in Euclidean spaces of dimensions 2, 3 and 5 are constructed. The corresponding classical and quantum mechanical analogues of the Kepler-Coulomb problem on these spheres are discussed. It is shown in Ref. [6], that oscillators on the sphere and the pseudosphere are related, by the so-called Bohlin transformation, with the Coulomb systems on the pseudosphere.

On the other hand, in the last decade an interest to such problems has grown abruptly due to experimental realization of nanostructures of different geometry [9-11]. The motion of particles on such surfaces should be described via quantum mechanics on curved spaces [12-15]. Besides single-electron states there were considered also two-electron states on spherical surfaces, in other words so-called spherical helium atoms were discussed [16-20].

The theoretical investigation of electron states in layered nanostructures is originated from the pioneering works of Chakraborty and Pietilainen [21-23]. Authors have considered one-electron and many-electron states in quantum rings at the presence of impurities,   as well as under the influence of a magnetic field. At the same time, taking into account that in the radial direction the movement of electron is restricted both on internal and external radiuses, Chakraborty and Pietilainen have suggested a model of confining potential having the form of a two-dimensional shifted oscillator
$$
V_{conf}(r)=\alpha(r-r_0)^2,
$$
where $\alpha$ characterizes the intensity of electron localization. Further there were studied optical, kinetic, spin, etc. properties of charge carriers localized in circinate and cylindrical layered nanostructures (See for example, [24-31]).

Nanolayers of spherical symmetry are being studied in Ref. [32-36]. The important peculiarity of spherical nanolayers is more flexible control of energy spectrum by changing both inner and outer radiuses (instead of only one outer radius as it is for the case of spherical quantum dot). Moreover, in particular cases, the results obtained for spherical nanolayer can be adapted for systems such as quantum well and spherical quantum dot. In Ref. [37-39] one and two electron states as well as optical properties of spherical nanolayers are investigated. In Refs [35], [36] it is assumed that small thicknesses of the spherical layer means that particle is localized on a spherical surface with some effective radius $R_1<R_{eff}<R_2$, where $R_1$ and $R_2$ are respectively the inner and outer radiuses of the layer.

In our recent work [36] we have examined electron states localized in a quantum ring on a spherical surface. As a confinement potential we have chosen the singular analog of the so-called $\mathbb{C}P^N$ -oscillator suggested by Bellucci and Nersessian in [40]
$$
V(\theta)=4\beta r_0^2{\tan}^2 \frac{\theta}{2}+\frac{\alpha}{4 r_0^2{\tan}^2 \frac{\theta}{2}},
$$
which is a spherical generalization of the Smorodinsky-Winternitz potential [41]. On the other hand, it is interesting to discuss the same angular problem of electron states in a quantum ring on a spherical surface (see Fig. 1) for the case when the angular confinement potential is chosen in the form of rectangular impenetrable walls.

\section{theory}

The Schr\"{o}dinger equation of the system can be written as follows
\begin{equation}
-\frac{\hbar^2}{2\mu}\left[ \frac{\partial^2}{\partial r^2}+\frac{2}{r}\frac{\partial}{\partial r}+\frac{1}{r^2}\nabla_{\theta,\varphi}^2 \right]\psi+\left( V_{conf}^{rad}(r)+V_{conf}^{ang}(\theta)\right) \psi=E\psi,
\end{equation}
where
\begin{equation}
V_{conf}^{rad}(r)=\begin{cases}
        0, &R_1<r<R_2,\vspace{10pt}\\
        \infty , &r<R_1, r>R_2,\\
    \end{cases}
\end{equation}
\begin{equation}
V_{ang}^{conf}(r)=\begin{cases}
        0, &\theta_1<\theta<\theta_2\vspace{10pt}\\
        \infty , &\theta<\theta_1, \theta>\theta_2\\
    \end{cases},
\end{equation}
and $\mu$ is the effective mass of electron. As it was mentioned above, the small thickness of the layer lets us assume that the electron will be in the ground state on the radial direction and move on the spherical surface of radius $r_0=R_{eff}=\left( R_1+R_2 \right)/2$ [35], [36]. Then for the radial wave function of the ground state we can write [35]:
\begin{equation}
R_0(r)=\sqrt{\frac{\pi \lambda}{2r}}\left( D_1 J_{\frac{1}{2}}(\lambda r)+D_2 J_{-\frac{1}{2}}(\lambda r)\right),
\end{equation}
where $\lambda=\sqrt{{2\mu E_0^{rad}}/{\hbar^2}}$, $J_{\nu}(x)$ is the Bessel function, $D_1$ and $D_2$ -- normalization constants. $R_0(r)$ satisfies the equation
\begin{equation}
-\frac{\hbar^2}{2\mu}\left( \frac{d^2}{dr^2}+\frac{2}{r}\frac{d}{dr}\right)R_0(r)+V_{conf}^{rad}(r)R_0(r)=E_0^{rad}R_0(r),
\end{equation}
where for $E_0^{rad}$ we have [35]:
\begin{equation}
E_0^{rad}=\frac{\pi^2\hbar^2N^2}{2\mu(R_2-R_1)^2},\qquad N=1,2,3,\ldots .
\end{equation}
Here $N$ is quantum number of the radial quantization.

Thus, taking into account the polar symmetry of the problem, we seek for the solution of the Schr\"{o}dinger equation in the form
\begin{equation}
\psi(r,\theta,\varphi)=R_0(r)e^{\imath m \varphi}P(\theta).
\end{equation}
Substituting (7) into (1) and introducing a notation
\begin{equation}
\frac{2\mu}{\hbar^2}r_0^2 (E-E_0^{rad})=\frac{2\mu}{\hbar^2}r_0^2 E_{ang}\equiv l(l+1),
\end{equation}
we obtain following equation:
\begin{equation}
\frac{d^2P}{d\theta^2}+\cot\theta\frac{dP}{d\theta}+\left(l(l+1)-\frac{m^2}{\sin^2\theta}\right)P(\theta)=0.
\end{equation}
Introducing a new variable
\begin{equation}
y\equiv sin^2 \frac\theta 2
\end{equation}
we arrive to equation
\begin{equation}
y(1-y)\frac{d^2P}{dy^2}+2\left(\frac{1}{2}-y\right)\frac{dP}{dy}+\frac{1}{4}\left( 4l(l+1) -\frac{m^2}{y}-\frac{m^2}{1-y}\right)P=0,
\end{equation}
solution of which is [42]
\begin{equation}
    \begin{aligned}
1)m\geq0& \\
&P(y)=C_1 y^{\frac m 2} (1-y)^{\frac m 2} {}_2 F_1  (m+l+1,m-l,1+m,y)\\
&+ C_2 y^{\frac m 2} (1-y)^{\frac m 2} {}_2 F_1  (m+l+1,m-l,1+m,1-y),\\
2)m<0&\\
&P(y)=C_1' y^{-\frac m 2} (1-y)^{\frac m 2} {}_2 F_1  (l+1,-l,1-m,y)\\
&+ C_2' y^{\frac m 2} (1-y)^{-\frac m 2} {}_2 F_1  (l+1,-l,1-m,1-y).\\
    \end{aligned}
\end{equation}
In original notations
\begin{equation}
    \begin{aligned}
1)m\geq0& \\
&P(\theta)=C_1 sin^m \frac{\theta}{2} cos^m \frac{\theta}{2} {}_2 F_1  \left(m+l+1,m-l,1+m,sin^2 \frac\theta 2\right)\\
&+ C_2 sin^m \frac{\theta}{2} cos^m \frac{\theta}{2} {}_2 F_1  \left(m+l+1,m-l,1+m,cos^2 \frac\theta 2\right),\\
2)m<0&\\
&P(\theta)=C_1' sin^{-m} \frac{\theta}{2} cos^m \frac{\theta}{2} {}_2 F_1  \left(l+1,-l,1-m,sin^2 \frac\theta 2\right)\\
&+ C_2' sin^m \frac{\theta}{2} cos^{-m} \frac{\theta}{2} {}_2 F_1  \left(l+1,-l,1-m,cos^2 \frac\theta 2\right).\\
    \end{aligned}
\end{equation}
For both cases a coefficient $C_2$ can be represented through $C_1$ via realization of the boundary condition $P(\theta_1)=0$. And $C_1$ is found from the normalization condition:
\begin{equation}
\int_{\theta_1}^{\theta_2}|P(\theta)|^2\sin\theta d\theta\equiv1.
\end{equation}

We can obtain the values of the quantum number $l$, describing the energy spectrum of the system, solving a transcendental equation
\begin{equation}
    \begin{vmatrix}
        P_1(\theta_1)&P_2(\theta_1)\\
        P_1(\theta_2)&P_2(\theta_2)\\
    \end{vmatrix}=0,
\end{equation}
where
\begin{equation}
    \begin{aligned}
1)m\geq0& \\
&P_1(\theta)=sin^m \frac{\theta}{2} cos^m \frac{\theta}{2} {}_2 F_1  \left(m+l+1,m-l,1+m,sin^2 \frac\theta 2\right)\\
&P_2(\theta)=sin^m \frac{\theta}{2} cos^m \frac{\theta}{2} {}_2 F_1  \left(m+l+1,m-l,1+m,cos^2 \frac\theta 2\right),\\
2)m<0&\\
&P_1(\theta)=sin^{-m} \frac{\theta}{2} cos^m \frac{\theta}{2} {}_2 F_1  \left(l+1,-l,1-m,sin^2 \frac\theta 2\right)\\
&P_2(\theta)=sin^m \frac{\theta}{2} cos^{-m} \frac{\theta}{2} {}_2 F_1  \left(l+1,-l,1-m,cos^2 \frac\theta 2\right).\\
    \end{aligned}
\end{equation}
Note that in contradistinction to the case of sphere in this system the quantum number $l$ is not an integer quantity.

The dependence of the ground state energy on the internal boundary angle for different values of the quantum number $m$ is shown in Fig. 2. Numerical calculations are performed for GaAs. The effective radius of the spherical surface is chosen as $r_0=100{\AA}$. The energy is represented in the terms of the hydrogen atom ground state energy: $\varepsilon=E_{ang}/\frac{m_0 e^4}{2\hbar^2}$. As we can see, with the increase of the boundary angle $\theta_1$ the energy increases due to the decrease of the electron localization area. The increase of the quantum number $m$ means the increase of the energy of the azimuthal motion, which leads to the increase of the total angular energy.

The increase of the boundary angle $\theta_2$ leads to the increase of the localization, and hence, to the decrease of the ground state energy (Fig. 3).

It should be noted, that in the case when $\theta_1\rightarrow 0, \theta_2\rightarrow \frac{\pi}{2}$, the problem is reduced to the problem of electron motion on the surface of semisphere [43].

\section{quantum transitions}

On the basis of the obtained results we can calculate the matrix elements of quantum transitions in the discussed system. If we consider electron transition from valence band into conductive band without taking into account excitonic effects, than for corresponding absorption coefficient we can use the formulae [44]
\begin{equation}
K=A\sum_{m,m',l,l'}\left|\int \psi^e_{m,l}\psi^h_{m',l'}dv\right|^2\delta(\hbar\omega-E_g-E^e_{l,m}-E^h_{l',m'}),
\end{equation}
where $A$ is a quantity proportional to the square of the matrix element taken by Bloch functions [45], $\omega$ is the frequency of the incident light, $E_g$ is the band gap of massive semiconductor. $\delta$ is the Dirac delta function, which provides the energy conservation law during the transitions. The selection rule for magnetic quantum number gives $m'=-m$. Then integration of radial and azimuthal parts of the wave functions gives 1, and finally for absorption coefficient we have
\begin{equation}
    \begin{split}
K=& A\sum_{l,l'}\int\limits_{\theta_1}^{\theta_2} \left[ C_1 sin^m \frac{\theta}{2} cos^m \frac{\theta}{2} {}_2 F_1  \left(m+l+1, m-l,1+m,sin^2 \frac\theta 2\right) \right.\\
&+\left. C_2 sin^m \frac{\theta}{2} cos^m \frac{\theta}{2} {}_2 F_1  \left(m+l+1,m-l,1+m,cos^2 \frac\theta 2\right) \right]\\
& \left[ C_1' sin^{-m} \frac{\theta}{2} cos^m \frac{\theta}{2} {}_2 F_1  \left(l+1,-l,1-m,sin^2 \frac\theta 2\right) \right.\\
&+ C_2' sin^m \frac{\theta}{2} cos^{-m} \frac{\theta}{2} \left. {}_2 F_1  \left(l+1,-l,1-m,cos^2 \frac\theta 2\right)\right] sin\theta d\theta\\
&\delta\left(\hbar\omega-E_g-\frac{\hbar^2 \pi^2}{(R_2-R_1)^2}\frac{\mu_h+\mu_e}{\mu_h\mu_e} -\frac{\hbar^2 l(l+1)}{2\mu_e r_0^2} -\frac{\hbar^2 l'(l'+1)}{2\mu_h r_0^2}\right).\\
    \end{split}
\end{equation}

Here we have chosen $m\geq0$.   Note that the imposition of boundary conditions on values of the polar angle $\theta (\theta_1\leq\theta\leq\theta_2)$ leads to the removal of the selection rules for the orbital quantum number $l$ due to violation of the spherical symmetry of the problem. Indeed, in case when $\theta$ varies between $[0,\pi]$ due to the orthogonality of the Legendre polynomials we have $l=l'$ [44].

The threshold frequency, starting from which an absorption takes place is determined from the conditions $m=0$, $l=l'=l_{ground { } state}$ (see Fig. 4), and has the following form:
\begin{equation}
\omega=\frac{E_g}{\hbar}+\frac{\hbar \pi^2}{\mu_{red}(R_2-R_1)^2} +\frac{\hbar l(l+1)}{2\mu_{red} r_0^2},
\end{equation}
where $\mu_{red}=\frac{\mu_h\mu_e}{\mu_h+\mu_e}$.

The size quantization energy is described by the difference of internal and external radiuses of the layer, while the geometrical parameter of the angular energy is the effective radius $r_0$. In Fig. 5 we have presented the dependence of the threshold frequency on the internal boundary angle $\theta_1$ for different values of the internal and external radiuses. As it might be expected, it repeats the ground state energy behavior, shown in Fig. 2. The increase of $r_0$ leads to the increase of the localization area, which respectively results in the decrease of the energy and threshold frequency.

\section{conclusion}

Thus, we have presented an analytical solution of the quantum problem of an electron on a spherical segment with angular confinement potential of the form of rectangular impenetrable walls. It is shown, that the solutions are expressed via combination of hypergeometric functions. It turned out that quantum number $l$, which is the analogue of orbital quantum number is not an integer quantity in this system. We have discussed the quantum transitions in this system and found that the selection rule for quantum number $l$ is removed due to the violation of spherical symmetry of the problem.

\begin{acknowledgements}
The authors are grateful to Professor George Pogosyan, Professor Armen Nersessian  and Dr. Vram Mughnetsyan for discussion of the results and valuable comments. The work was performed within the framework of the State basic program "Investigation of physical properties of quantum nanostructures with a complex geometry and different confining potentials".
\end{acknowledgements}

\newpage \includegraphics[width=17cm]{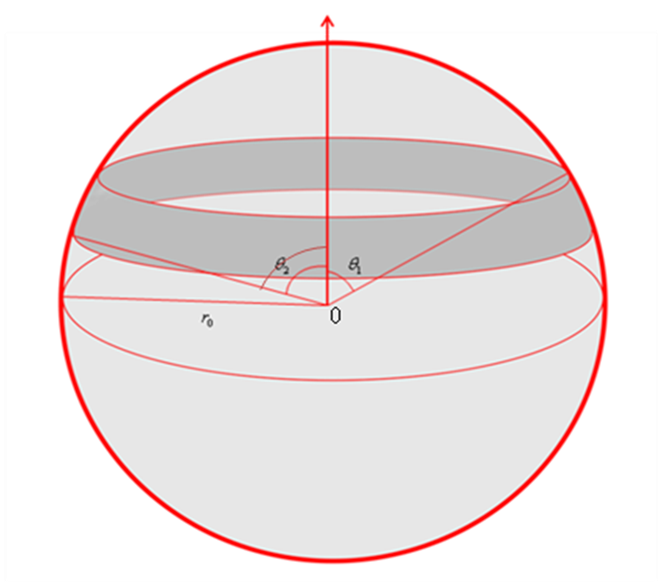} \\ \vfill \center{\textbf{Fig. 1}. Spherical segment}

\newpage \includegraphics[width=17cm]{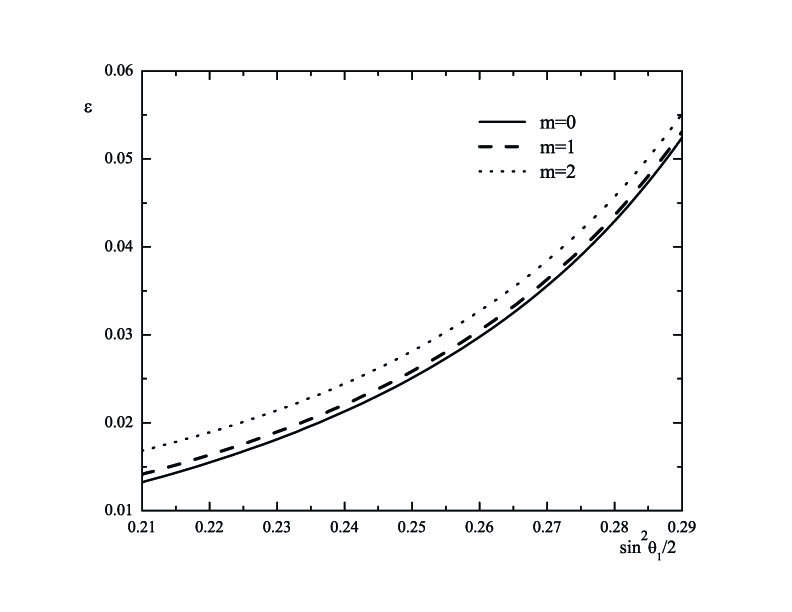} \\ \vfill \center{\textbf{Fig. 2}. The ground state energy dependence on the internal boundary angle $\theta_1$}

\newpage \includegraphics[width=17cm]{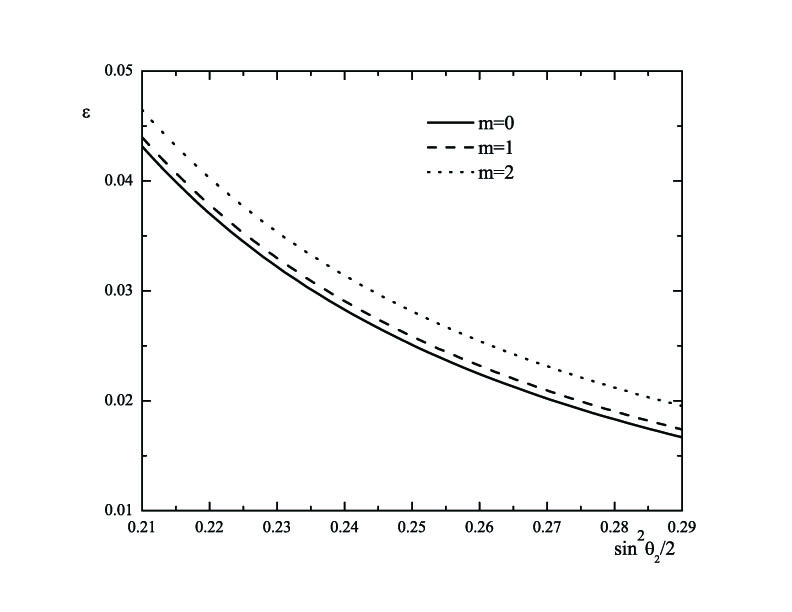} \\ \vfill \center{\textbf{Fig. 3}. The ground state energy dependence on the external boundary angle $\theta_2$}

\newpage \includegraphics[width=14cm]{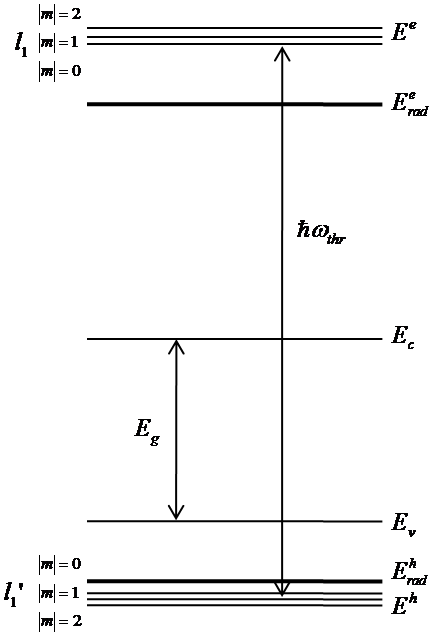} \\ \vfill \center{\textbf{Fig. 4}. Energetic diagram of quantum transitions}

\newpage \includegraphics[width=17cm]{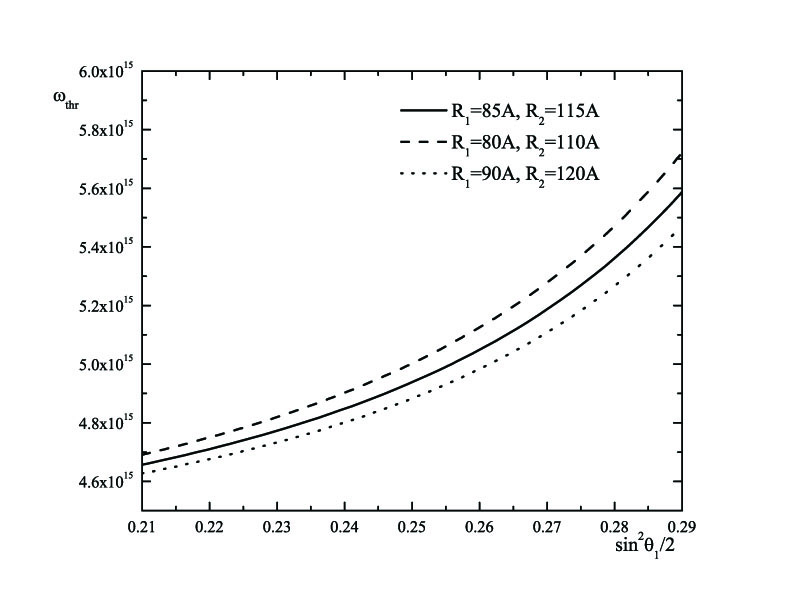} \\ \vfill \center{\textbf{Fig. 5}. The threshold frequency dependence on the internal boundary angle $\theta_1$}

\end{document}